\documentclass[11pt,a4paper]{article}

\usepackage{amsmath}
\usepackage{amssymb}
\usepackage{graphicx}
\graphicspath{{./}}
\usepackage{url}

\makeatletter
\newcommand{\affiliation}[1]{\gdef\@affiliation{#1}}
\newcommand{\emailAdd}[1]{\gdef\@email{#1}}
\renewcommand{\abstract}[1]{\gdef\@abstract{#1}}
\newcommand{\keywords}[1]{\gdef\@keywords{#1}}
\gdef\@affiliation{}
\gdef\@email{}
\gdef\@abstract{}
\gdef\@keywords{}
\renewcommand{\maketitle}{%
  \begin{center}
    {\Large\bfseries \@title\par}
    \vspace{0.8em}
    {\@author\par}
    \vspace{0.5em}
    {\@affiliation\par}
    \vspace{0.5em}
    {\ttfamily \@email\par}
  \end{center}
  \vspace{1em}
  \noindent\textbf{Abstract.} \@abstract\par
  \vspace{0.8em}
  \noindent\textbf{Keywords:} \@keywords\par
  \vspace{1.2em}
}
\makeatother

\newcommand{\safeincludegraphics}[2][]{%
  \IfFileExists{#2}{%
    \includegraphics[#1]{#2}%
  }{%
    \fbox{%
      \begin{minipage}[c][0.22\textheight][c]{0.78\textwidth}
        \centering
        Missing figure file:\\[0.5em]
        \texttt{\detokenize{#2}}
      \end{minipage}%
    }%
  }%
}

\title{\boldmath Finite-Time Bounds on Conformal Vacuum Energy Relaxation in Scalar-Tensor Gravity}

\author{Yousef Bisabr}

\affiliation{Department of Physics, Shahid Rajaee Teacher Training University,\\
Lavizan, Tehran 16788, Iran}

\emailAdd{y-bisabr@sru.ac.ir}

\abstract{
We investigate finite time vacuum energy relaxation in conformally
coupled scalar-tensor cosmology  distinguishing the suppression
accumulated along a physical trajectory from its asymptotic rate.
For a canonical scalar with an exponential potential and constant
conformal coupling, the autonomous system exhibits a stability exchange
between scalar dominated and vacuum kinetic scaling solutions. In the
power-law metric $f(R)=f_0R^n$ realization with $n>2$, the stable
scalar dominated branch has a sustained relaxation rate below $0.579$
decades per Einstein frame e-fold. We additionally derive a finite time
upper bound that accounts for depletion of the vacuum relative to the
scalar potential. For the specified reference initial state
$\Omega_{\Lambda,i}=0.99$ and $\Omega_{{\rm kin},i}=10^{-4}$, the bound
excludes $60$ decades of suppression within $60$ e-folds throughout the
$n>2$ family: $D_\Lambda(60)<47.7$. Numerical trajectories illustrate
why this bound is conservative and why the transient contribution cannot
be inferred from the final attractor alone. For the reference
$f(R)\propto R^{2.14}$ solution, $D_\Lambda(60)\simeq7.53$ decades
including a transient excess of about $3.27$ decades relative to
uniform evolution at the asymptotic rate. The relaxed quantity is the
dimensionless vacuum to gravitational scale hierarchy, its evolution
also entails variation of the conformal frame Planck mass. The bounds
are conditional on the homogeneous model and specified initial state,
not a demonstration of complete cosmological viability.
}

\keywords{vacuum energy; vacuum energy relaxation;
scalar-tensor cosmology; dynamical systems;
$f(R)$ gravity; Brans-Dicke theory}

\begin{document}

\maketitle
\flushbottom
\section{Introduction}

The disparity between the observed cosmological vacuum scale and the
natural scales suggested by quantum field theory remains a central
problem in cosmology \cite{Weinberg1989}-\cite{Martin2012}. Scalar-tensor
and modified gravity approaches can alter the gravitational role of a
vacuum component through additional scalar degrees of freedom
\cite{Clifton2012}-\cite{bis3}. In a conformally coupled description,
a vacuum density that is constant in the matter frame appears as
$\rho_\Lambda=A^4\tilde\rho_\Lambda$ in the Einstein frame. Its
suppression is therefore tied to the actual scalar displacement and
to the evolution of the effective gravitational scale
\cite{bis2,Damour1992}.\\
The phase space dynamics of exponential scalar potentials, coupled
quintessence and scalar-tensor cosmologies are well established
\cite{Copeland1998,Amendola2000,Bahamonde2018,Faraoni2005,Holden2000}.
The existence of a relaxation channel or a stable attractor does not,
however, establish how much vacuum suppression can occur within a
finite cosmological interval. The asymptotic rate can underestimate
the contribution from early kinetic motion whereas the unrestricted
kinematic capacity can substantially overestimate what the equations
of motion permit. The distinction is particularly consequential when
asking whether a prescribed hierarchy can be generated within a
specified number of e-folds.\\
We examine this finite time question in a homogeneous Einstein frame
model containing a canonical scalar, an exponential potential and a
conformally coupled vacuum component. The autonomous formulation
identifies the stable relaxation branches and their rates. We then
specialize to power-law metric $f(R)=f_0R^n$ gravity whose Einstein frame
potential and universal conformal coupling are fixed by $n$
\cite{Sotiriou2010,DeFelice2010,Alho2016}. For $n>2$ the stable
scalar dominated branch obeys a model specific asymptotic rate ceiling.
Our additional result is an initial condition dependent finite time
bound obtained by combining the scalar evolution with the depletion of
the vacuum relative to its potential. For the reference initial state,
it excludes a $60$-decade reduction in $60$ Einstein frame e-folds for
the entire $n>2$ family. This statement is conditional: it does not
extend to arbitrary initial states, coupling functions or cosmological
histories.\\
Direct integrations are used to compare the analytical restriction with
the realized suppression and to distinguish the suppression accumulated
before settling from the excess above the asymptotic rate contribution.
Brans-Dicke (BD) and field dependent coupling examples illustrate the
model dependence of the relaxation history. We also clarify that the
frame independent dimensionless vacuum to Planck hierarchy changes
through a corresponding evolution of the matter frame gravitational
scale. The present work concerns homogeneous dynamics; perturbative
stability, local gravity constraints and a complete radiation-matter
history remain outside its scope.\\
The paper is organized as follows. Section~2 introduces the
dynamical framework, the vacuum energy relaxation diagnostic
and its conformal frame interpretation. Section~3 develops
the autonomous system, identifies its critical points and
examines their stability and asymptotic relaxation rates.
Section~4 specializes the analysis to power-law metric
$f(R)$ gravity and derives both the asymptotic relaxation
ceiling and the initial condition dependent finite time
bound. Section~5 presents numerical solutions for the
$f(R)$, BD and field dependent coupling models
examines the transient contribution and compares the
realized suppression with the analytical bounds.
Section~6 discusses the physical interpretation and
limitations of the results including the associated
evolution of the gravitational scale. Finally,
Section~7 summarizes the main conclusions.

~~~~~~~~~~~~~~~~~~~~~~~~~~~~~~~~~~~~~~~~~~~~~~~~~~~~~~~~~~~~~~~~~~~~~~~~~~~~~~~~~~~~~~~~~~~~~~~~~~~~
\section{Dynamical framework and relaxation diagnostic}

We consider a scalar-tensor theory in its Einstein frame
representation, described by the action \cite{bis2,bis3} \cite{Damour1992}
\begin{equation}
S =
\int d^4x\,\sqrt{-g}
\left[
\frac{M_{\rm Pl}^2}{2}R
-\frac{1}{2}g^{\mu\nu}\partial_\mu\varphi\partial_\nu\varphi
-V(\varphi)
\right]
+S_\Lambda[A^2(\varphi)g_{\mu\nu}],
\label{action}
\end{equation}
where $\varphi$ is a canonical scalar field, $V(\varphi)$ is its
potential and the vacuum sector is coupled to the conformally related
metric
\begin{equation}
\tilde g_{\mu\nu}=A^2(\varphi)g_{\mu\nu}.
\end{equation}
The strength of the conformal interaction is characterized by
\begin{equation}
\beta(\varphi)
\equiv
M_{\rm Pl}\frac{d\ln A(\varphi)}{d\varphi}.
\end{equation}
The vacuum component is taken to have a constant energy density
$\tilde{\rho}_\Lambda$ with respect to the conformal metric
$\tilde g_{\mu\nu}$ \cite{Damour1992}. Its action may therefore be written as
\[
S_\Lambda
=
-\int d^4x\,\sqrt{-\tilde g}\,\tilde{\rho}_\Lambda
=
-\int d^4x\,\sqrt{-g}\,
A^4(\varphi)\tilde{\rho}_\Lambda,
\]
where we have used
$\sqrt{-\tilde g}=A^4(\varphi)\sqrt{-g}$. Accordingly, the quantity
entering the Einstein frame background equations is
\[
\rho_\Lambda(\varphi)
=
A^4(\varphi)\tilde{\rho}_\Lambda.
\]
Thus $\rho_\Lambda$ should be understood as the Einstein-frame image
of a vacuum component that is constant with respect to the conformal
metric rather than as an independently conserved Einstein frame
cosmological constant.\\
The corresponding dimensionless hierarchy is frame independent:
if $M_{{\rm Pl},J}=M_{\rm Pl}/A$ denotes the coefficient-defined
Planck mass in the conformal frame, then
$\tilde\rho_\Lambda/M_{{\rm Pl},J}^4
=\rho_\Lambda/M_{\rm Pl}^4$. Thus the suppression studied here
corresponds to an evolving gravitational scale in the conformal frame;
it is not a decrease of $\tilde\rho_\Lambda$ itself.\\
We assume a spatially flat FLRW background and use the e-fold variable
$N\equiv\ln a$ where $a$ is the scale factor. Taking a logarithmic
derivative of the above relation with respect to $N$ and using the
definition of $\beta(\varphi)$ gives
\begin{equation}
\frac{d\ln\rho_\Lambda}{dN}
=
4\beta(\varphi)
\frac{1}{M_{\rm Pl}}
\frac{d\varphi}{dN}.
\label{vacuum_exchange}
\end{equation}
The factor of four therefore follows directly from the conformal
scaling of the four-dimensional volume element and represents the
energy exchange between the scalar field and the vacuum component in
the Einstein frame.
To quantify the resulting relaxation, we define the accumulated
vacuum energy suppression in decades by
\begin{equation}
D_\Lambda(N)
=
\log_{10}\!\left[
\frac{\rho_{\Lambda,i}}{\rho_\Lambda(N)}
\right].
\label{eq:Dlambda}
\end{equation}
We also introduce the scalar kinetic fraction
\begin{equation}
\Omega_{\rm kin}
\equiv
\frac{\dot{\varphi}^{\,2}}
{6M_{\rm Pl}^2H^2},
\label{eq:Omegakin}
\end{equation}
where $H\equiv\dot a/a$. Since $dN=Hdt$, it follows that
\begin{equation}
\frac{1}{M_{\rm Pl}}
\left|\frac{d\varphi}{dN}\right|
=
\sqrt{6\Omega_{\rm kin}}.
\label{eq:fieldvelocity}
\end{equation}
Combining this relation with the vacuum exchange law gives the
kinematic capacity relation
\begin{equation}
D_\Lambda
\leq
\frac{4\sqrt{6}}{\ln 10}
\int |\beta(\varphi)|
\sqrt{\Omega_{\rm kin}}\,dN .
\label{eq:paper2_capacity}
\end{equation}
Equation (8) will be used only as a diagnostic. It specifies the maximum
vacuum energy suppression available along a prescribed coupling-weighted
scalar trajectory but does not determine whether such a trajectory is
dynamically realized. The problem studied here is therefore to determine,
once the scalar potential and conformal coupling are specified, what
trajectory $\Omega_{\rm kin}(N)$ is generated by the field equations, which
asymptotic attractor is selected, and at what rate vacuum-energy relaxation
is accumulated.
~~~~~~~~~~~~~~~~~~~~~~~~~~~~~~~~~~~~~~~~~~~~~~~~~~~~~~~~~~~~~~~~~~~~~~~~~~~~~~~~~~~~~~~~~~~~~~~~~~~~
\section{Autonomous dynamics and asymptotic relaxation}
\label{sec:numerics}

We now formulate the homogeneous equations as an autonomous system,
identify the physical critical points and determine the relaxation rate
selected by each stable branch. The fixed point classification follows
the established scalar field dynamical systems framework; its role here
is to distinguish the sustained rate from finite time suppression.

\subsection{Autonomous formulation and physical phase space} \label{subsec:num_system}

We define the dimensionless canonical field and its e-fold velocity by
\begin{equation}
 \varphi\equiv\frac{\phi}{M_{\rm Pl}},\qquad
 x\equiv\frac{d\varphi}{dN},\qquad
 \Omega_{\rm kin}=\frac{x^2}{6}.
 \label{eq:num_variables}
\end{equation}
The scalar potential is taken to be
\begin{equation}
 V(\varphi)=V_0e^{-\lambda\varphi},
 \label{eq:num_potential}
\end{equation}
with constant slope $\lambda$. We retain the canonical scalar and the
conformally coupled vacuum component and introduce
\begin{equation}
 \Omega_\Lambda\equiv
 \frac{\rho_\Lambda}{3M_{\rm Pl}^2H^2},
 \qquad
 \Omega_V\equiv
 \frac{V}{3M_{\rm Pl}^2H^2}.
\end{equation}
The Friedmann constraint is then
\begin{equation}
 1=\Omega_{\rm kin}+\Omega_V+\Omega_\Lambda
   =\frac{x^2}{6}+\Omega_V+\Omega_\Lambda,
 \label{eq:num_constraint}
\end{equation}
while the Raychaudhuri equation becomes
\begin{equation}
 \frac{H'}{H}=-\frac{x^2}{2},
 \label{eq:num_Hprime}
\end{equation}
where a prime denotes $d/dN$.
Combining the scalar equation of motion with the vacuum exchange equation
gives
\begin{align}
 \varphi'&=x,
 \label{eq:num_phiprime}\\
 x'&=-\left(3-\frac{x^2}{2}\right)x
      +3\lambda\Omega_V
      -12\beta(\varphi)\Omega_\Lambda,
 \label{eq:num_xprime}\\
 \Omega_\Lambda'&=
 \Omega_\Lambda\left[4\beta(\varphi)x+x^2\right],
 \label{eq:num_OmLprime}
\end{align}
with
\begin{equation}
 \Omega_V=1-\frac{x^2}{6}-\Omega_\Lambda.
 \label{eq:num_OmV}
\end{equation}
The physical phase space is fixed by positivity of the three fractional
energy densities,
\begin{equation}
 \Omega_\Lambda\geq0,\qquad
 \Omega_V\geq0,\qquad
 \Omega_{\rm kin}\geq0,
\end{equation}
which implies
\begin{equation}
 -\sqrt6\leq x\leq\sqrt6,
 \qquad
 0\leq\Omega_\Lambda\leq1-\frac{x^2}{6}.
 \label{eq:num_physical_region}
\end{equation}
Equation~\eqref{eq:num_Hprime} also gives the effective equation of state 
\begin{equation}
 w_{\rm eff}
 =-1-\frac{2}{3}\frac{H'}{H}
 =-1+\frac{x^2}{3}.
 \label{eq:num_weff}
\end{equation}
The logarithmic vacuum suppression defined in Eq.~\eqref{eq:Dlambda}
evolves according to
\begin{equation}
D_\Lambda'
\equiv
\frac{dD_\Lambda}{dN}
=
-\frac{4}{\ln 10}\,\beta(\varphi)x ,
\label{eq:num_Dprime}
\end{equation}
This quantity represents the actual relaxation rate generated by the
dynamical evolution of the scalar field. By contrast, the kinematic
capacity introduced in Section~2 corresponds to the maximum rate
allowed by a prescribed scalar trajectory and is given by
\begin{equation}
\left(D_\Lambda^{\rm cap}\right)'
=
\frac{4}{\ln 10}\,
|\beta(\varphi)x|.
\label{eq:num_Dcapprime}
\end{equation}
The capacity rate in Eq.~\eqref{eq:num_Dcapprime} provides a
kinematic upper bound on the vacuum relaxation rate. For a monotonic
relaxation trajectory satisfying $\beta x<0$, this bound is saturated
so that
\begin{equation}
D_\Lambda=D_\Lambda^{\rm cap},
\end{equation}
up to the common initial normalization. This equality is not an
independent dynamical result but follows directly from the vacuum
exchange law. The nontrivial dynamical information is instead contained in the trajectory
$x(N)$ or equivalently $\Omega_{\rm kin}(N)$ which determines the rate at
which vacuum energy suppression is accumulated.

\subsection{Critical points and stability}
\label{subsec:critical_points}

We now determine the asymptotic structure of the constant coupling system.
When $\beta$ is constant, $\phi$ does not appear explicitly in the equations
for $x$ and $\Omega_\Lambda$. Eliminating $\Omega_V$ with Eq.~(17) gives the
two-dimensional autonomous system
\begin{align}
 x'={}&-\left(3-\frac{x^2}{2}\right)x
 +3\lambda\left(1-\frac{x^2}{6}-\Omega_\Lambda\right)
 -12\beta\Omega_\Lambda,
 \label{eq:dyn_x}\\
 \Omega_\Lambda'={}&
 \Omega_\Lambda x(x+4\beta).
 \label{eq:dyn_OmL}
\end{align}
The critical points satisfy $x'=\Omega_\Lambda'=0$ and are
\begin{align}
 P_\pm&=(\pm\sqrt6,0),
 \label{eq:crit_kin}\\
 P_\varphi&=(\lambda,0),
 \label{eq:crit_phi}\\
 P_{\rm dS}&=
 \left(0,\frac{\lambda}{\lambda+4\beta}\right),
 \label{eq:crit_ds}\\
 P_\beta&=
 \left(-4\beta,1-\frac{8}{3}\beta^2\right).
 \label{eq:crit_beta}
\end{align}
The points $P_\pm$ are kinetic dominated solutions with
$\Omega_{\rm kin}=1$ and $w_{\rm eff}=1$. At $P_\varphi$,
\begin{equation}
 \Omega_{\rm kin}=\frac{\lambda^2}{6},
 \qquad
 \Omega_V=1-\frac{\lambda^2}{6},
 \qquad
 \Omega_\Lambda=0,
 \label{eq:Pphi_densities}
\end{equation}
so this point belongs to the physical phase space for $\lambda^2\leq6$.
Its effective equation of state is
\begin{equation}
 w_{\rm eff}=-1+\frac{\lambda^2}{3}.
\end{equation}
The point $P_{\rm dS}$ has $x=0$ and therefore $w_{\rm eff}=-1$.
Requiring both $\Omega_\Lambda$ and $\Omega_V$ to be non-negative gives
away from the degenerate case $\lambda+4\beta=0$, the condition
$\beta\lambda>0$. 
Finally, $P_\beta$ is a vacuum kinetic scaling solution with
\begin{equation}
 \Omega_{\rm kin}=\frac{8}{3}\beta^2,
 \qquad
 \Omega_\Lambda=1-\frac{8}{3}\beta^2,
 \qquad
 \Omega_V=0,
 \label{eq:Pbeta_densities}
\end{equation}
and therefore exists in the physical region when
\begin{equation}
 \beta^2\leq\frac{3}{8}.
 \label{eq:Pbeta_existence}
\end{equation}
\begin{table}[t]
\centering
\caption{Critical points of the constant coupling autonomous system
and their existence conditions.}
\label{tab:critical_points}
\begin{tabular}{c c c c c}
\hline
Point & $x_*$ & $\Omega_{\Lambda,*}$ &
$\Omega_{{\rm kin},*}$ & Existence \\
\hline
$P_\pm$
& $\pm\sqrt{6}$
& $0$
& $1$
& Always
\\[1mm]

$P_\varphi$
& $\lambda$
& $0$
& $\lambda^2/6$
& $\lambda^2\leq6$
\\[1mm]

$P_{\rm dS}$
& $0$
& $\dfrac{\lambda}{\lambda+4\beta}$
& $0$
& $\beta\lambda>0$
\\[3mm]

$P_\beta$
& $-4\beta$
& $1-\dfrac{8}{3}\beta^2$
& $\dfrac{8}{3}\beta^2$
& $\beta^2\leq\dfrac{3}{8}$
\\
\hline
\end{tabular}
\end{table}
Linearizing Eqs.~\eqref{eq:dyn_x}-\eqref{eq:dyn_OmL} about a critical
point gives
\begin{equation}
 {\cal J}=
 \begin{pmatrix}
 -3+\dfrac{3}{2}x^2-\lambda x
 &
 -3\lambda-12\beta
 \\[2mm]
 \Omega_\Lambda(2x+4\beta)
 &
 x(x+4\beta)
 \end{pmatrix}.
 \label{eq:dyn_jacobian}
\end{equation}
At $P_\varphi$ the eigenvalues are
\begin{equation}
 \mu_1=-3+\frac{\lambda^2}{2},
 \qquad
 \mu_2=\lambda(\lambda+4\beta),
 \label{eq:eigen_phi}
\end{equation}
so that $P_\varphi$ is a stable attractor when
\begin{equation}
 \lambda^2<6,
 \qquad
 \lambda(\lambda+4\beta)<0.
 \label{eq:stability_phi}
\end{equation}
For the kinetic points,
\begin{align}
 P_+:\qquad&
 \mu_1=6-\sqrt6\,\lambda,
 &
 \mu_2=6+4\sqrt6\,\beta,
 \label{eq:eigen_plus}\\
 P_-:\qquad&
 \mu_1=6+\sqrt6\,\lambda,
 &
 \mu_2=6-4\sqrt6\,\beta.
 \label{eq:eigen_minus}
\end{align}
At $P_{\rm dS}$,
\begin{equation}
 \mu_\pm=
 \frac{-3\pm\sqrt{9-48\beta\lambda}}{2},
 \label{eq:eigen_ds}
\end{equation}
and therefore, on its generic physical branch $\beta\lambda>0$,
$P_{\rm dS}$ is stable. It is a stable node for
$0<\beta\lambda<3/16$ and a stable spiral for
$\beta\lambda>3/16$. 
At $P_\beta$,
\begin{equation}
 \mu_1=8\beta^2-3,
 \qquad
 \mu_2=4\beta(4\beta+\lambda).
 \label{eq:eigen_beta}
\end{equation}
For the sign choice relevant to the relaxation solutions considered below,
$\beta<0$ and $\lambda>0$, the de Sitter point lies outside the physical
phase space. Provided the corresponding fixed points exist, the remaining
attractor structure reduces to
\begin{equation}
 \lambda<4|\beta|
 \quad\Longrightarrow\quad
 P_\varphi\ {\rm stable},
 \label{eq:attractor_phi}
\end{equation}
whereas provided the vacuum kinetic scaling point lies in the
physical phase space,
\begin{equation}
 \lambda>4|\beta|,
 \qquad
 \beta^2<\frac{3}{8}
 \quad\Longrightarrow\quad
 P_\beta\ {\rm stable}.
 \label{eq:attractor_beta}
\end{equation}
The stability conditions alone do not determine whether the corresponding
attractor describes accelerated expansion. At $P_\varphi$, the equation of
state given above implies acceleration only when $\lambda^2<2$. At
$P_\beta$ using $x=-4\beta$ and
$w_{\rm eff}=-1+x^2/3$ gives
\[
 w_{\rm eff}^{(\beta)}
 =
 -1+\frac{16\beta^2}{3},
\]
so the vacuum kinetic scaling solution accelerates only when
$\beta^2<1/8$. Thus dynamical stability and sustained vacuum energy
relaxation do not by themselves guarantee an accelerating asymptotic
cosmology.
\begin{table}[!ht]
\centering
\caption{Stable asymptotic solutions and vacuum-energy relaxation rates
for the relaxation branch $\beta<0$ and $\lambda>0$.}
\label{tab:relaxation_attractors}
\begin{tabular}{c c c c}
\hline
Attractor & Stability condition & $x_*$ & $D'_{\Lambda,*}$ \\
\hline

$P_\varphi$
&
$\lambda<4|\beta|,\ \lambda^2<6$
&
$\lambda$
&
$\displaystyle\frac{4|\beta|\lambda}{\ln 10}$
\\[3mm]

$P_\beta$
&
$\lambda>4|\beta|,\ \beta^2<\frac{3}{8}$
&
$4|\beta|$
&
$\displaystyle\frac{16\beta^2}{\ln 10}$
\\

\hline
\end{tabular}
\end{table}
Thus
\begin{equation}
 \lambda_c=4|\beta|
 \label{eq:lambda_c}
\end{equation}
marks the transition between the scalar field dominated and
vacuum kinetic scaling attractors. At $\lambda=\lambda_c$ the two
branches merge and the system possesses the continuous equilibrium line
\begin{equation}
 x=-4\beta=\lambda,
 \qquad
 0\leq\Omega_\Lambda
 \leq1-\frac{8}{3}\beta^2.
 \label{eq:critical_line}
\end{equation}
The fixed-point structure has a direct interpretation in terms of vacuum
relaxation. From Eq.~\eqref{eq:num_Dprime}, the asymptotic relaxation rate
at $P_\varphi$ is
\begin{equation}
 \left.D_\Lambda'\right|_{P_\varphi}
 =-\frac{4\beta\lambda}{\ln10},
 \label{eq:D_phi_rate}
\end{equation}
while at $P_\beta$,
\begin{equation}
 \left.D_\Lambda'\right|_{P_\beta}
 =\frac{16\beta^2}{\ln10}.
 \label{eq:D_beta_rate}
\end{equation}
At $P_{\rm dS}$ the scalar is stationary and $D_\Lambda'=0$.
For the relaxation branch $\beta<0$, $\lambda>0$, Eqs.~\eqref{eq:D_phi_rate}
and \eqref{eq:D_beta_rate} may therefore be written as
\begin{equation}
 D'_{\Lambda,*}=
 \begin{cases}
 \displaystyle \frac{4|\beta|\lambda}{\ln10},
 & \lambda<4|\beta|,\\[2mm]
 \displaystyle \frac{16\beta^2}{\ln10},
 & \lambda>4|\beta|,
 \end{cases}
 \label{eq:D_asym_piecewise}
\end{equation}
whenever the corresponding attractor exists. At the stability exchange
$\lambda=\lambda_c=4|\beta|$, the two expressions coincide,
\begin{equation}
 \left.D'_{\Lambda,*}\right|_{\lambda_c}
 =\frac{16\beta^2}{\ln10}.
 \label{eq:D_rate_continuity}
\end{equation}
Thus the bifurcation changes the composition and equat on of state of the
stable asymptotic solution without producing a discontinuous jump in the
vacuum relaxation rate. The fixed point analysis therefore determines not
only the asymptotic cosmological state but also the rate at which relaxation
capacity is accumulated.
\section{Power-law $f(R)$ realization}
\label{subsec:powerlaw_fr}

The exponential potential used above has a direct realization in metric
$f(R)$ gravity. Consider
\begin{equation}
 f(R)=f_0R^n.
 \label{eq:powerlaw_fr}
\end{equation}
Under the standard transformation from metric $f(R)$ gravity to its
Einstein frame scalar-tensor representation
\cite{Sotiriou2010,DeFelice2010,Alho2016}, we define
\begin{equation}
 F(R)\equiv\frac{df}{dR},
 \qquad
 \varphi=\sqrt{\frac32}\ln F,
 \label{eq:fr_scalar}
\end{equation}
and restrict throughout to the branch $F(R)>0$, for which the conformal
transformation to the Einstein frame is well defined and the effective
gravitational coupling has the standard sign. The Einstein frame potential is
\begin{equation} 
 V(\varphi)
 =
 \frac{M_{\rm Pl}^2}{2}\frac{RF-f}{F^2}.
 \label{eq:fr_EF_potential}
\end{equation}
For $f(R)=f_0R^n$, one has $F=nf_0R^{n-1}$ and the potential becomes
\begin{equation}
 V(\varphi)=V_0e^{-\lambda_n\varphi},
\end{equation}
where
\begin{equation}
 \lambda_n=
 \sqrt{\frac23}\frac{n-2}{n-1}.
 \label{eq:lambda_powerlaw}
\end{equation}
Thus the constant coupling exponential system contains power-law metric
$f(R)$ gravity as the particular realization
\begin{equation}
 \beta=-\frac{1}{\sqrt6},
 \qquad
 \lambda=\lambda_n.
 \label{eq:fr_identification}
\end{equation}
The reference value $\lambda=0.1$ used below corresponds to
\begin{equation}
 n=
 \frac{2-\sqrt{3/2}\,\lambda}
      {1-\sqrt{3/2}\,\lambda}
 \simeq2.14.
 \label{eq:n_reference}
\end{equation}
Accordingly, the metric $f(R)$ reference integration may be interpreted
explicitly as a representative trajectory of the power-law model
$f(R)\propto R^{2.14}$.
The power-law parameter also determines which attractor branch is available.
For $n>2$,
\begin{equation}
 0<\lambda_n<\sqrt{\frac23},
 \label{eq:lambda_n_range}
\end{equation}
whereas for metric $f(R)$ gravity
\begin{equation}
 \lambda_c=4|\beta|
 =\frac{4}{\sqrt6}
 =2\sqrt{\frac23}.
 \label{eq:lambda_fr_critical}
\end{equation}
Hence
\begin{equation}
 \lambda_n<\lambda_c
 \qquad (n>2),
 \label{eq:fr_attractor_condition}
\end{equation}
so the entire $n>2$ power-law branch lies on the $P_\varphi$-attractor
side of the dynamical transition. In particular, varying $n$ within this
branch cannot move the system to the vacuum-kinetic scaling attractor
$P_\beta$. This is a model specific consequence of the general phase space
analysis.\\
The same mapping yields a stronger dynamical restriction on sustained
relaxation than the absolute kinematic ceiling. Substituting
$|\beta|=1/\sqrt6$ and Eq.~\eqref{eq:lambda_powerlaw} into
Eq.~\eqref{eq:D_phi_rate} gives, for $n>2$,
\begin{equation}
 D'_{\Lambda,*}(n)
 =\frac{4}{3\ln10}\frac{n-2}{n-1}
 <\frac{4}{3\ln10}
 \simeq0.579
 \qquad {\rm decades/e\!-\!fold}.
 \label{eq:fr_dynamical_ceiling}
\end{equation}
Consequently, if the evolution is governed by the asymptotic power-law
attractor, suppression by $60$ and $120$ decades requires respectively
\begin{equation}
 \Delta N_{60}> \frac{3\ln10}{4}(60)\simeq103.6,
 \qquad
 \Delta N_{120}> \frac{3\ln10}{4}(120)\simeq207.2.
 \label{eq:fr_efold_floor}
\end{equation}
The limiting rate in Eq.~\eqref{eq:fr_dynamical_ceiling} is only one third
of the absolute metric $f(R)$ kinematic ceiling obtained by formally
maintaining $\Omega_{\rm kin}=1$. This distinction makes explicit the
difference between kinematically available capacity and the smaller rate
that can be sustained by this particular attractor family. 
Here the power-law model is used as an explicit analytical realization of
the constant coupling exponential system. No claim is made that a pure
$f(R)=f_0R^n$ form provides a complete phenomenologically viable
description over the full cosmological history (in particular, see \cite{bis5,Amendola2007}).

\subsection{A conditional finite time bound}
\label{subsec:fr_finite}
The asymptotic ceiling does not bound the suppression accumulated during
the approach to the attractor. A finite-time restriction follows directly
from the autonomous equations. Put $b=|\beta|$, $c=4b$ and
$r_i=\rho_{\Lambda,i}/V_i$. On the relaxation branch
$0<\lambda<c<\sqrt6$, trajectories with $0\leq x_i\leq c$ remain in
$0\leq x\leq c$. Defining
$q=\rho_\Lambda/(\rho_\Lambda+V)$, one finds
\begin{equation}
 x'=3\left(1-\frac{x^2}{6}\right)
 [\lambda+(c-\lambda)q-x],\qquad
 \frac{r}{r_i}=e^{-(c-\lambda)(\phi-\phi_i)},
 \label{eq:fr_q_dynamics}
\end{equation}
where $r=\rho_\Lambda/V$ and $\phi=\varphi/M_{\rm Pl}$.
Integrating the scalar equation and using
$0\leq q\leq1$ and $0\leq x\leq c$ yields
\begin{equation}
 \Delta D_\Lambda\leq\frac{c}{\ln10}
 \min\!\left\{c\Delta N,
 \frac{c^2\Delta N+\ln(1+r_i)+
 c\sqrt{2/3}\,\operatorname{artanh}(x_i/\sqrt6)}
 {2c-\lambda}\right\}.
 \label{eq:fr_finite_bound}
\end{equation}
The bound follows from the exact integral
$(c-\lambda)\int xq\,dN
=\ln[(1+r_i)/(1+r_f)]\leq\ln(1+r_i)$;
the last term in Eq.~\eqref{eq:fr_finite_bound} conservatively retains
the initial-velocity contribution. It therefore applies to the entire
$n>2$ family for the stated initial-velocity range, including slopes
smaller than $x_i$.\\
For metric $f(R)$, $c=4/\sqrt6$ and
$\lambda_n<\sqrt{2/3}$. At the reference initial state used in
Section~5, $\Omega_{\Lambda,i}=0.99$,
$\Omega_{{\rm kin},i}=10^{-4}$ and $r_i=100$.
The right-hand side of Eq.~\eqref{eq:fr_finite_bound} is largest as
$n\to\infty$. Consequently, over $\Delta N=60$,
\begin{equation}
 D_\Lambda(60)<47.7\quad\text{decades}\qquad(n>2),
 \label{eq:fr_finite_ceiling}
\end{equation}
which excludes $60$ decades within $60$ Einstein frame e-folds for
this initial state. Inverting the same inequality gives the necessary
condition $\Delta N\gtrsim76$ for that target. These are conditional
finite time statements, not universal bounds on $f(R)$ cosmology or
predictions that the bound can be saturated.

\section{Phase space trajectories and numerical realization}
\label{subsec:phase_portraits}

\begin{figure*}[t]
\centering
\safeincludegraphics[width=0.98\textwidth]{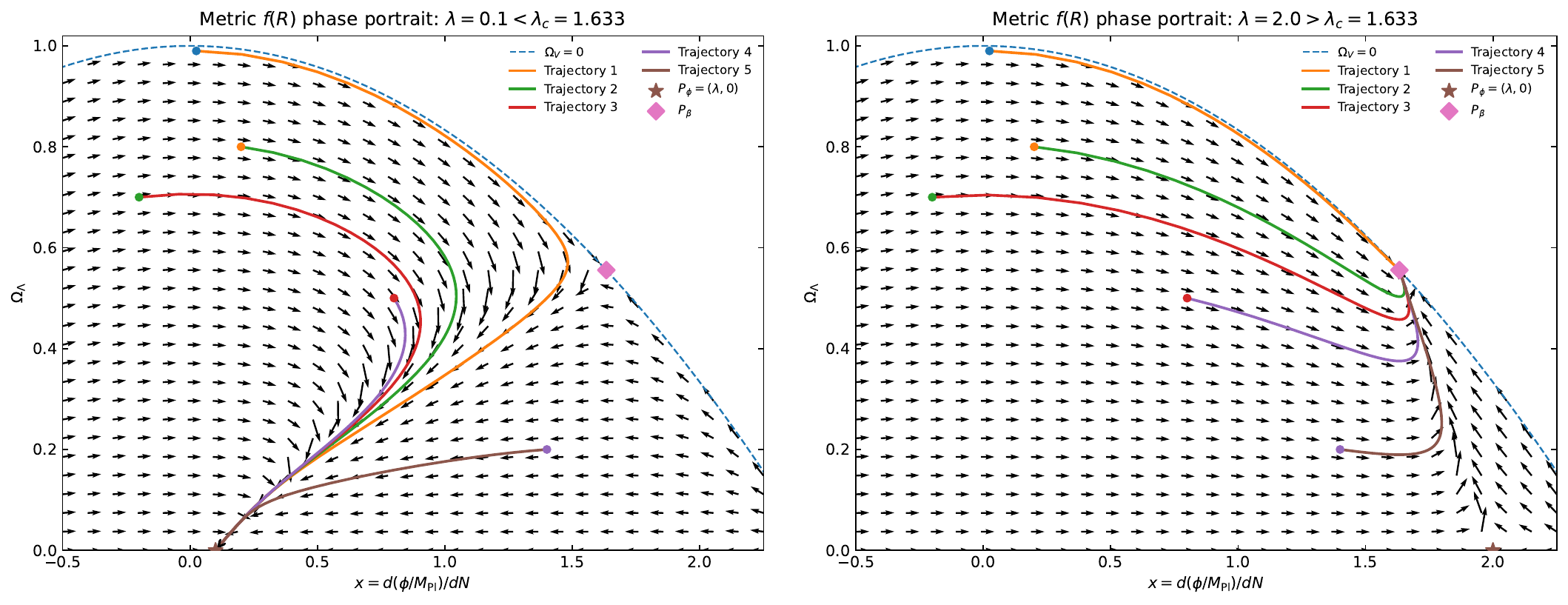}
\caption{Phase space flow in the dynamically relevant portion of the
physical region for the constant coupling exponential system with
$\beta=-1/\sqrt6$. (a) $\lambda=0.1<\lambda_c=4/\sqrt6$ corresponding
to the power-law realization $f(R)\propto R^{2.14}$ for which the
scalar field dominated point $P_\varphi$ is the stable attractor.
(b) $\lambda=2>\lambda_c$ illustrating the vacuum kinetic scaling
regime of the general constant coupling exponential system in which the
stable attractor is $P_\beta$. The dashed curve denotes the boundary
$\Omega_V=0$ of the physical phase space.}
\label{fig:phase_portraits}
\end{figure*}

The fixed point structure can be visualized directly in the
$(x,\Omega_\Lambda)$ plane. We first consider the power-law metric $f(R)$
reference model used in the numerical analysis below,
\begin{equation}
 \beta=-\frac{1}{\sqrt6},
 \qquad
 \lambda=0.1,
 \qquad
 n\simeq2.14.
 \label{eq:phase_reference}
\end{equation}
Since
\begin{equation}
 \lambda<4|\beta|,
\end{equation}
the scalar field dominated point
\begin{equation}
 P_\varphi=(0.1,0)
\end{equation}
is the stable late-time attractor. 
The left panel of Fig.~\ref{fig:phase_portraits} shows the flow in the
dynamically relevant portion of the physical phase space,
\begin{equation}
 0\leq\Omega_\Lambda\leq1-\frac{x^2}{6}.
\end{equation}
Trajectories starting from different admissible initial conditions converge
toward $P_\varphi$. During the approach they may undergo a substantial
transient excursion in the kinetic direction anticipating the transient
enhancement of $\Omega_{\rm kin}$ found in the direct integrations below.
The phase portrait therefore gives a geometric interpretation of the
relaxation process: a significant amount of vacuum suppression can be
accumulated during the transient evolution even though the final attractor
has a relatively small kinetic fraction.\\
For comparison, the right panel of Fig.~\ref{fig:phase_portraits} keeps the
same constant coupling, $\beta=-1/\sqrt6$, but takes
$\lambda=2>\lambda_c$. This panel is included to illustrate the attractor
transition of the general constant coupling exponential system, not as a
preferred power-law $f(R)$ realization. In this regime $P_\varphi$ is
unstable and the trajectories approach
\begin{equation}
 P_\beta=
 \left(
 \frac{4}{\sqrt6},\frac{5}{9}
 \right).
\end{equation}

\subsection{Reference models and common initial state}
\label{subsec:num_models}

We now turn to the direct numerical integrations. Equations
\eqref{eq:num_phiprime}-\eqref{eq:num_OmLprime} are solved with an
adaptive Runge--Kutta algorithm using relative and absolute tolerances
$3\times10^{-10}$ and $10^{-12}$, respectively. The Friedmann constraint
is monitored through Eq.~\eqref{eq:num_OmV}; all solutions quoted below
remain inside the physical region
$\Omega_\Lambda\geq0$, $\Omega_V\geq0$, and
$\Omega_{\rm kin}\geq0$ over the displayed interval.\\
We compare three representative couplings over $\Delta N=60$ using the
common initial state
\begin{equation}
 \Omega_{\Lambda,i}=0.99,\qquad
 \Omega_{{\rm kin},i}=10^{-4},\qquad
 \varphi_i=0,
 \label{eq:num_initial}
\end{equation}
with $\lambda=0.1$. Thus
$x_i=\sqrt{6\times10^{-4}}$ and $\Omega_{V,i}=0.0099$.
These choices define a common vacuum dominated starting point and are not
intended as observationally preferred parameters.\\
The three reference models are
\begin{align}
 {\rm (i)}\quad&
 \beta=-\frac{1}{\sqrt6},
 &&
 \text{power-law }f(R),\quad n\simeq2.14,
 \label{eq:num_case_fr}\\
 {\rm (ii)}\quad&
 \beta=-\frac{1}{\sqrt{46}},
 &&
 \text{BD with }\omega_{\rm BD}=10,
 \label{eq:num_case_bd}\\
 {\rm (iii)}\quad&
 \beta(\varphi)=-0.2e^{0.1\varphi},
 &&
 \text{field-dependent scalar--tensor coupling}.
 \label{eq:num_case_var}
\end{align}
Since the dimensionless scalar variable used in the autonomous system
satisfies $\varphi=\phi/M_{\rm Pl}$, the conformal coupling defined in
Section 2 may equivalently be written as
$\beta(\varphi)=d\ln A/d\varphi$. The field-dependent coupling in
Eq.~\eqref{eq:num_case_var} therefore corresponds to
\[
 A(\varphi)
 =
 A_0\exp\!\left[-2e^{0.1\varphi}\right],
\]
where $A_0$ is an arbitrary normalization constant. Thus the chosen
field-dependent coupling is generated by a well-defined conformal factor
rather than being introduced independently at the level of the dynamical
equations.\\
This functional form is chosen as a simple illustrative example in which
the magnitude of the conformal coupling grows monotonically along the
relevant scalar trajectory. The purpose is not to single out this
parameterization phenomenologically but to isolate the dynamical effect
of an evolving coupling. Qualitatively similar enhancement may arise when
$|\beta(\varphi)|$ grows sufficiently along the relaxation trajectory
although the realized relaxation also depends on the accompanying scalar
kinetic history.\\
For comparison among models with constant and field dependent couplings,
we define the kinetic-weighted effective coupling over the integration
interval by
\begin{equation}
\beta_{\rm eff}\equiv
\frac{\displaystyle\int_{N_i}^{N_f}
|\beta(\phi)|\sqrt{\Omega_{\rm kin}}\,dN}
{\displaystyle\int_{N_i}^{N_f}
\sqrt{\Omega_{\rm kin}}\,dN}.
\end{equation}
For a constant coupling this quantity reduces to $|\beta|$ whereas for a
field-dependent coupling it characterizes the coupling sampled by the
kinetic history of the scalar field.\\
For the power-law $f(R)$ reference model, $\lambda=0.1$ is well below
$\lambda_c=4/\sqrt6\simeq1.633$. The fixed point analysis therefore predicts
the asymptotic state
\begin{equation}
 P_\varphi=(0.1,0),
 \qquad
 \Omega_{{\rm kin},*}
 =\frac{0.1^2}{6}
 \simeq1.67\times10^{-3},
 \label{eq:fr_reference_attractor}
\end{equation}
which provides a direct analytical benchmark for the numerical solution.

\subsection{Reference evolution: kinetic history and accumulated relaxation}
\label{subsec:num_reference}

Table 3 summarizes the reference integrations and Fig.~2 shows the
corresponding accumulated vacuum energy suppression.
\begin{table}[t]
\centering
\caption{Reference numerical integrations over $\Delta N=60$ for the
common initial state in Eq.~\eqref{eq:num_initial}. The final vacuum
suppression is given in decades.}
\label{tab:numerical_results}
\begin{tabular}{lcccc}
\hline\hline
Model & $\beta_{\rm eff}$ & $D_\Lambda(60)$
& $\Omega_{\rm kin}^{\rm max}$ & $\Omega_{\Lambda,f}$ \\
\hline
power-law $f(R)$ ($n\simeq2.14$)
 & $0.408$ & $7.53$ & $0.366$ & $8.61\times10^{-6}$ \\
BD, $\omega_{\rm BD}=10$
 & $0.147$ & $4.24$ & $0.0563$ & $2.90\times10^{-2}$ \\
variable $\beta(\varphi)$
 & $0.394$ & $8.43$ & $0.156$ & $1.28\times10^{-6}$ \\
\hline\hline
\end{tabular}
\end{table}
For the power-law $f(R)$ reference model, approximately $7.5$ decades are
generated over the displayed interval. The $\omega_{\rm BD}=10$ model
reaches only about $4.2$ decades because its fixed coupling is substantially
weaker. By contrast, the field dependent model reaches about $8.4$ decades. Although it starts from $|\beta_i|=0.2$, the magnitude of the coupling grows
along the trajectory and compensates for its smaller kinetic fraction.
\begin{figure}[t]
\centering
\safeincludegraphics[width=0.78\textwidth]{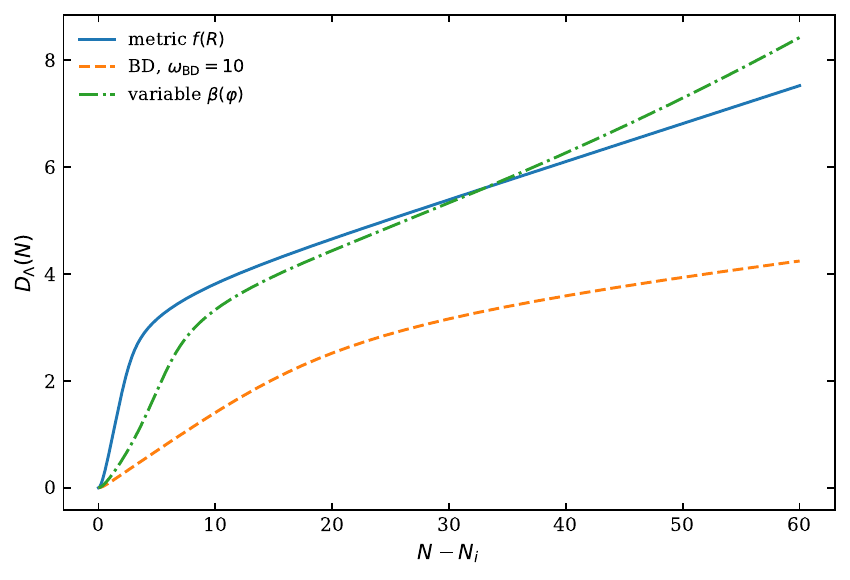}
\caption{Accumulated vacuum suppression $D_\Lambda(N)$ for the three
reference models over $\Delta N=60$. All trajectories start from the
same vacuum dominated state, Eq.~\eqref{eq:num_initial} with
$\lambda=0.1$. The metric-$f(R)$ trajectory corresponds to the
power-law realization $f(R)\propto R^{2.14}$.}
\label{fig:num_relaxation}
\end{figure}
The corresponding kinetic histories are shown in
Fig.~\ref{fig:num_kinetic}. They demonstrate why replacing
$\Omega_{\rm kin}(N)$ by its maximum value throughout the full interval can
substantially overestimate the available capacity. In the power-law $f(R)$
case, $\Omega_{\rm kin}$ reaches approximately $0.37$, but this enhancement
is transient. The relevant quantity is the integrated
coupling-weighted history $|\beta|\sqrt{\Omega_{\rm kin}}$, not the largest
instantaneous kinetic fraction.
\begin{figure}[t]
\centering
\safeincludegraphics[width=0.78\textwidth]{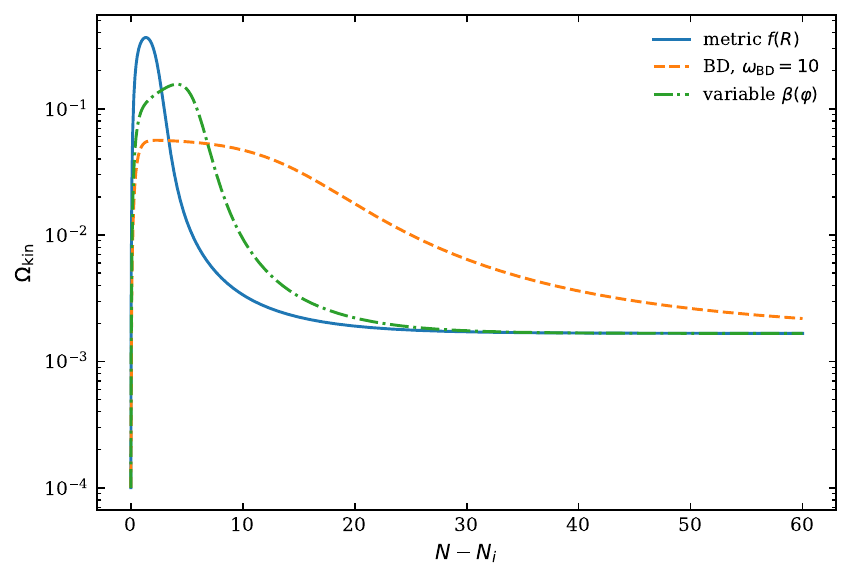}
\caption{Scalar kinetic fraction $\Omega_{\rm kin}=x^2/6$ for the
reference trajectories of Fig.~\ref{fig:num_relaxation}. The transient
kinetic enhancement illustrates why the integrated relaxation budget is
more informative than $\Omega_{\rm kin}^{\rm max}$ alone.}
\label{fig:num_kinetic}
\end{figure}
For all three reference solutions, $\beta x<0$ throughout the plotted
interval. Consequently, $D_\Lambda$ and $D_\Lambda^{\rm cap}$ agree to
numerical precision providing a direct implementation check of the exchange
equation. The differences among the three curves are therefore dynamical:
the coupling and scalar equation determine the kinetic history available to
accumulate the kinematic budget.\\
The late-time behavior of the power-law $f(R)$ model also agrees with the
fixed point prediction. After the transient kinetic enhancement, the
trajectory approaches $P_\varphi$, with
$x\rightarrow\lambda=0.1$ and
$\Omega_{\rm kin}\rightarrow\lambda^2/6$. The numerical evolution therefore
provides a direct realization of the attractor structure derived above.\\
The importance of the pre-asymptotic evolution can be quantified with a
simple settling diagnostic. Defining the trajectory to be near the
$P_\varphi$ attractor once
$|x-\lambda|/\lambda<5\%$ and this condition remains satisfied thereafter,
the reference metric $f(R)$ solution settles at approximately
$N-N_i\simeq21.9$. By then it has accumulated
$D_\Lambda\simeq4.80$ decades, compared with
$D_\Lambda(60)\simeq7.53$. Thus about $64\%$ of the total suppression in
this reference integration is accumulated before the trajectory has settled
within this neighborhood of its final attractor. The remaining
$\simeq2.72$ decades are consistent with the fixed point rate
$D'_{\Lambda,*}=4|\beta|\lambda/\ln10\simeq0.0709$ over the subsequent
evolution. The precise numerical fraction depends on the adopted settling
criterion but the conclusion that the transient contributes materially to
the integrated relaxation is insensitive to this convention.
A complementary, settling criterion independent measure is the
transient excess above evolution at the final attractor rate:
\begin{equation}
 D_\Lambda(N)=\frac{4b\lambda}{\ln10}\Delta N
 +\frac{4b}{\ln10}\int_{N_i}^{N}[x(\bar N)-\lambda]d\bar N.
 \label{eq:transient_excess}
\end{equation}
For the reference $f(R)$ trajectory, the asymptotic rate contribution
over $60$ e-folds is $4.26$ decades and the transient excess is
approximately $3.27$ decades, or $43.5\%$ of the total. This differs
from the $64\%$ accumulated before the chosen settling time: the latter
also includes suppression that would occur at the asymptotic rate.

\subsection{Dependence on the initial kinetic fraction}
\label{subsec:num_kin_scan}

\begin{figure}[t]
\centering
\safeincludegraphics[width=0.78\textwidth]{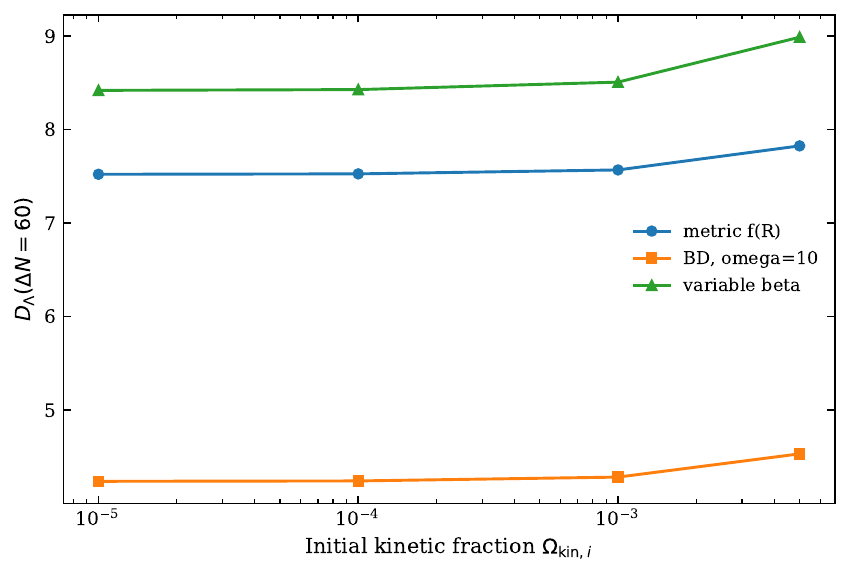}
\caption{Dependence of the accumulated relaxation after $60$ e-folds on
the initial kinetic fraction. The weak sensitivity at small
$\Omega_{{\rm kin},i}$ shows that the subsequent dynamically generated
kinetic history rather than the initial kinetic energy alone controls the
integrated relaxation.}
\label{fig:num_kin_scan}
\end{figure}

To test whether the reference results depend sensitively on the chosen
initial scalar motion, we repeat the integrations for
\begin{equation}
 \Omega_{{\rm kin},i}
 =10^{-5},\;10^{-4},\;10^{-3},\;5\times10^{-3},
 \label{eq:num_kin_scan}
\end{equation}
while keeping $\Omega_{\Lambda,i}=0.99$, $\varphi_i=0$, and
$\lambda=0.1$. The upper value is chosen so that the initial potential
fraction remains positive. The results are given in
Fig.~\ref{fig:num_kin_scan} and Table~\ref{tab:kin_scan}.
\begin{table}[t]
\centering
\caption{Final relaxation $D_\Lambda(60)$ as the initial scalar kinetic
fraction is varied.}
\label{tab:kin_scan}
\begin{tabular}{ccccc}
\hline\hline
$\Omega_{{\rm kin},i}$
& $10^{-5}$ & $10^{-4}$ & $10^{-3}$ & $5\times10^{-3}$ \\
\hline
power-law $f(R)$ ($n\simeq2.14$)
& 7.52 & 7.53 & 7.57 & 7.82 \\
BD, $\omega_{\rm BD}=10$
& 4.24 & 4.24 & 4.28 & 4.53 \\
variable $\beta(\varphi)$
& 8.42 & 8.43 & 8.51 & 8.99 \\
\hline\hline
\end{tabular}
\end{table}
The scan shows that increasing the initial kinetic fraction does enhance the
final suppression but only weakly in the small kinetic energy regime. The
system rapidly develops a coupling- and potential dependent kinetic history
and therefore loses much of its memory of the precise initial value. For
example, increasing $\Omega_{{\rm kin},i}$ by two orders of magnitude from
$10^{-5}$ to $10^{-3}$ changes the power-law $f(R)$ result only from
$7.52$ to $7.57$ decades. A more noticeable enhancement occurs only when
the initial kinetic fraction is large enough to modify the early trajectory.
This result sharpens the interpretation of the analytic bound: a large
instantaneous or initial $\Omega_{\rm kin}$ is not sufficient by itself;
what matters is a sufficiently sustained coupling-weighted kinetic history.
\subsection{BD parameter scan}
\label{subsec:num_bd_scan}

\begin{figure}[t]
\centering
\safeincludegraphics[width=0.78\textwidth]{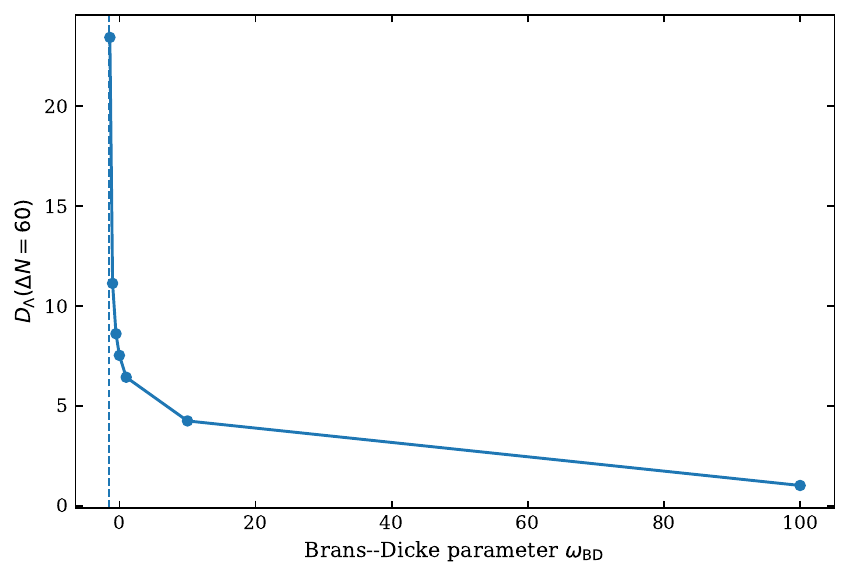}
\caption{Numerical vacuum suppression after $60$ e-folds as a function of
the BD parameter. The dashed vertical line marks the formal
boundary $\omega_{\rm BD}=-3/2$. The rapid enhancement toward this
boundary reflects the growth of $|\beta_{\rm BD}|$ and should not be
interpreted as evidence that the near-boundary regime is phenomenologically
viable.}
\label{fig:num_bd_scan}
\end{figure}

For constant BD theory the Einstein frame coupling is
\begin{equation}
 |\beta_{\rm BD}|=
 \frac{1}{\sqrt{4\omega_{\rm BD}+6}},
 \label{eq:num_beta_bd}
\end{equation}
which decreases as $\omega_{\rm BD}$ increases and grows as the ghost-free
boundary $\omega_{\rm BD}=-3/2$ is approached from above. To examine the
dynamical consequence, we integrate the same initial state for
\begin{equation}
 \omega_{\rm BD}=-1.4,-1,-0.5,0,1,10,100.
\end{equation}
The resulting relaxation is listed in Table~\ref{tab:bd_scan} and shown in
Fig.~\ref{fig:num_bd_scan}.
\begin{table}[t]
\centering
\caption{Dependence of the numerical relaxation on the constant
BD parameter for $\Delta N=60$ and the reference initial state.}
\label{tab:bd_scan}
\begin{tabular}{ccc}
\hline\hline
$\omega_{\rm BD}$ & $|\beta_{\rm BD}|$ & $D_\Lambda(60)$ \\
\hline
$-1.4$ & 1.581 & 23.44 \\
$-1.0$ & 0.707 & 11.13 \\
$-0.5$ & 0.500 & 8.61 \\
$0$ & 0.408 & 7.53 \\
$1$ & 0.316 & 6.43 \\
$10$ & 0.147 & 4.24 \\
$100$ & 0.0496 & 1.01 \\
\hline\hline
\end{tabular}
\end{table}
The numerical trend is fully consistent with the analytic capacity
argument. Metric $f(R)$, corresponding to $\omega_{\rm BD}=0$ at the level
of the conformal coupling, lies between the enhanced negative-
$\omega_{\rm BD}$ solutions and the progressively weaker positive-
$\omega_{\rm BD}$ cases. At $\omega_{\rm BD}=100$, only about one decade is
relaxed during the reference interval whereas $\omega_{\rm BD}=-1.4$
produces more than $23$ decades. The latter trajectory also develops a large
kinetic fraction, reaching
$\Omega_{\rm kin}^{\rm max}\simeq0.92$, which illustrates the dynamical
price of the enhanced relaxation. Values close to $-3/2$ should therefore
be regarded here as diagnostic tests of the capacity formula rather than as
preferred cosmological models.\\
The dynamical system analysis further clarifies this trend. For constant
BD coupling, the stability exchange
$\lambda=4|\beta_{\rm BD}|$ occurs at
\begin{equation}
 \omega_{\rm BD}^{(c)}
 =\frac{4}{\lambda^2}-\frac{3}{2}.
 \label{eq:bd_critical_omega}
\end{equation}
For the reference slope $\lambda=0.1$, this gives
$\omega_{\rm BD}^{(c)}=398.5$. Hence every model in
Table~\ref{tab:bd_scan} including $\omega_{\rm BD}=100$ remains on the
$P_\varphi$ side of the bifurcation. The increase of $D_\Lambda(60)$ toward
$\omega_{\rm BD}\rightarrow-3/2^+$ therefore occurs within a single
attractor class. It reflects the growth of the direct coupling together
with the coupling-dependent kinetic history rather than a transition
between distinct late-time fixed points.

\subsection{Summary of numerical implications}
\label{subsec:num_implications}

The combined dynamical and numerical analysis leads to several conclusions.
First, the upper bound is most naturally interpreted as a relaxation budget:
the scalar dynamics determines how much of the kinematically allowed
coupling-weighted path is actually accumulated. The fixed point analysis
sharpens this interpretation by relating the asymptotic cosmological state
directly to the asymptotic relaxation rate.\\
Second, the exponential potential provides a useful bridge between the
general scalar-tensor system and an explicit metric $f(R)$ realization.
For $f(R)=f_0R^n$, the slope is fixed by
Eq.~\eqref{eq:lambda_powerlaw}; the reference choice $\lambda=0.1$
corresponds to $n\simeq2.14$. On the $n>2$ branch,
$\lambda_n<\lambda_c$, so the entire branch remains on the scalar field
dominated side of the attractor transition. Moreover,
Eq.~\eqref{eq:fr_dynamical_ceiling} gives the model specific asymptotic
ceiling $D'_{\Lambda,*}<0.579$ decades per e-fold substantially below the
absolute kinematic capacity. The reference numerical trajectory also shows
that transient motion can materially enhance the integrated suppression
above the rate inferred from the final attractor alone.\\
Third, in the reference integrations considered here, positive constant
BD parameters reduce both the direct coupling and the dynamically
generated kinetic activity relative to the metric $f(R)$ case, making large
relaxation progressively more difficult. A field dependent scalar-tensor coupling provides an additional
degree of freedom: in the example considered here, the evolving coupling
produces a larger suppression than the fixed metric $f(R)$ case even though
its maximum kinetic fraction is smaller.\\
None of the moderate reference models, however, approaches a
$60$-$120$ decade hierarchy within $60$ e-folds. Even the strongly coupled
$\omega_{\rm BD}=-1.4$ diagnostic case reaches only about $23$ decades and
does so with a kinetic fraction approaching unity. The numerical solutions
therefore reinforce the central kinematic message: very large vacuum
relaxation over a finite interval requires a large integrated budget
\begin{equation}
 {\cal B}
 =\int|\beta|\sqrt{\Omega_{\rm kin}}\,dN,
\end{equation}
which must be supplied by a sufficiently strong coupling, sustained scalar
motion, a sufficiently long relaxation epoch or some combination of these
ingredients. The distinction between this kinematic requirement and the
dynamical viability of any particular realization is central to the
discussion that follows.

~~~~~~~~~~~~~~~~~~~~~~~~~~~~~~~~~~~~~~~~~~~~~~~~~~~~~~~~~~~~~~~~~~~~~~~~~~~~~~~~~~~~~~~~~~~~~~~~~~~~~~~~~~~~~~~~~~

\section{Discussion: finite-time versus asymptotic relaxation}

The fixed point analysis determines the sustained rate but does not
constrain the full transient by itself. The power-law $f(R)$ realization
makes this distinction explicit: its $n>2$ branch obeys the asymptotic
ceiling in Eq.~\eqref{eq:fr_dynamical_ceiling} whereas
Eq.~\eqref{eq:fr_finite_bound} limits the integrated suppression during
a specified finite interval. For the reference initial state, the
family-wide ceiling of Eq.~\eqref{eq:fr_finite_ceiling} excludes a
$60$-decade target in $60$ Einstein-frame e-folds. The numerical
reference model achieves $7.53$ decades, showing that the analytical
upper bound is conservative rather than a prediction of realized
suppression.\\
The transient also requires a precise definition. The $64\%$ of
suppression accumulated before the reference solution settles within
$5\%$ of its final scalar velocity measures an early-time fraction.
Equation~\eqref{eq:transient_excess} instead isolates the additional
$3.27$ decades above uniform evolution at the asymptotic rate. Both
measures show why a late-time attractor alone cannot determine a
finite time relaxation budget. The Brans--Dicke scan and the
field-dependent example further demonstrate the dependence on the
coupling-weighted kinetic history; strongly coupled near-boundary
trajectories are diagnostics, not established viable cosmologies.\\
The physical interpretation also matters. Although
$\tilde\rho_\Lambda$ is constant in the conformal frame, the
dimensionless ratio $\tilde\rho_\Lambda/M_{{\rm Pl},J}^4$ decreases
in both frame descriptions. Exactly,
\begin{equation}
 D_\Lambda=4\log_{10}
 \left(\frac{M_{{\rm Pl},J,f}}{M_{{\rm Pl},J,i}}\right).
 \label{eq:frame_tradeoff}
\end{equation}
Thus $60$ decades of suppression would entail a $15$-decade increase
in the coefficient-defined Jordan frame Planck mass. This is not,
without accounting for scalar-mediated forces, a direct prediction
for the locally measured Newton constant. Nor does reducing
$\rho_\Lambda$ ensure that the independent scalar potential or total
effective dark-energy contribution becomes small.\\
Our conclusions concern homogeneous dynamics under specified
assumptions. Perturbative stability, local gravity bounds and a
consistent radiation-matter history are not addressed. These
requirements, together with the gravitational-scale evolution, must
be satisfied before the mechanism can be regarded as a complete
cosmological solution.

~~~~~~~~~~~~~~~~~~~~~~~~~~~~~~~~~~~~~~~~~~~~~~~~~~~~~~~~~~~~~~~~~~~~~~~~~~~~~~~~~~~~~~~~~~~~
\section{Conclusions}

We have examined the finite-time suppression of a conformally coupled
vacuum component in homogeneous scalar-tensor cosmology. For a constant
coupling and exponential potential, the established fixed-point
structure selects the asymptotic relaxation rate with a continuous
rate across the stability exchange at $\lambda=4|\beta|$.
The $n>2$ power-law metric $f(R)$ branch remains scalar dominated and
obeys the sustained rate ceiling $D'_{\Lambda,*}<0.579$ decades per
e-fold.\\
Combining scalar dynamics with depletion of the vacuum relative to its
potential gives an additional finite-time bound. For the reference
initial state, it implies $D_\Lambda(60)<47.7$ decades for every
$n>2$, excluding $60$ decades within $60$ Einstein frame e-folds under
those assumptions. The reference $f(R)\propto R^{2.14}$ trajectory
instead yields $7.53$ decades including $3.27$ decades of transient
excess above the asymptotic-rate contribution. BD and
field dependent coupling examples illustrate the dependence on the
full coupling-weighted scalar history.\\
The suppression describes a changing dimensionless vacuum to Planck
hierarchy and necessarily accompanies variation of the conformal-frame
gravitational scale. It does not establish suppression of the entire
scalar plus vacuum energy sector or full phenomenological viability.
The finite-time bound is conditional on the model and initial state;
perturbative and observational tests remain necessary.\\

\paragraph{Acknowledgement.}
This work is supported by Shahid Rajaee Teacher Training University under
grant No.~23905/380001.

~~~~~~~~~~~~~~~~~~~~~~~~~~~~~~~~~~~~~~~~~~~~~~~~~~~~~~~~~~~~~~~~~~~~~~~~~~~~~~~~~~~~~~~~~~~~~~~~~~~~~~~~~~~~~~~~~~~~~~~~~~~~~~~~~~~


\begin{thebibliography}{99}

\bibitem{Weinberg1989}
S.~Weinberg,
\textit{The cosmological constant problem},
Rev.\ Mod.\ Phys.\ \textbf{61} (1989) 1.

\bibitem{Carroll2001}
S.M.~Carroll,
\textit{The cosmological constant},
Living Rev.\ Relativ.\ \textbf{4} (2001) 1
[astro-ph/0004075].

\bibitem{Padmanabhan2003}
T.~Padmanabhan,
\textit{Cosmological constant---the weight of the vacuum},
Phys.\ Rept.\ \textbf{380} (2003) 235--320
[hep-th/0212290].

\bibitem{Martin2012}
J.~Martin,
\textit{Everything you always wanted to know about the cosmological constant
problem (but were afraid to ask)},
C.\ R.\ Physique \textbf{13} (2012) 566--665
[arXiv:1205.3365 [astro-ph.CO]].

\bibitem{Clifton2012}
T.~Clifton et al.,
\textit{Modified gravity and cosmology},
Phys.\ Rept.\ \textbf{513} (2012) 1
[arXiv:1106.2476 [astro-ph.CO]].

\bibitem{bis1}Y. Bisabr, \emph{Mechanism for a Decaying Cosmological Constant}, Class.Quant.Grav. \textbf{19} (2002) 2369-2374 [arXiv:gr-qc/0109087]

\bibitem{bis2}Y. Bisabr, \emph{Gravitational Coupling, Dynamical Changes of Units and The Cosmological Constant Problem}, Comm. Theo. Phys. \textbf{72}, 035403 (2020) [arXiv:1904.09336]

\bibitem{bis3}Y. Bisabr, \emph{Cosmological Constant in Chameleon Brans-Dicke Theory},  Int. J. Mod. Phys. D \textbf{28}, 1950022 (2019) [arXiv:1711.05426] 

\bibitem{Copeland1998}
E.J.~Copeland, A.R.~Liddle and D.~Wands,
\textit{Exponential potentials and cosmological scaling solutions},
Phys.\ Rev.\ D \textbf{57} (1998) 4686
[gr-qc/9711068].

\bibitem{Amendola2000}
L.~Amendola,
\textit{Coupled quintessence},
Phys.\ Rev.\ D \textbf{62} (2000) 043511
[astro-ph/9908023].

\bibitem{Bahamonde2018}
S.~Bahamonde, C.~G.~B\"ohmer, S.~Carloni, E.~J.~Copeland,
W.~Fang and N.~Tamanini,
Dynamical systems applied to cosmology: dark energy and modified gravity,
Phys. Rept. \textbf{775--777}, 1 (2018)
[arXiv:1712.03107 [gr-qc]].

\bibitem{Faraoni2005}
V.~Faraoni,
\textit{Phase space geometry in scalar--tensor cosmology},
Annals Phys.\ \textbf{317} (2005) 366--382
[gr-qc/0502015].

\bibitem{Holden2000}
D.J.~Holden and D.~Wands,
\textit{Self-similar cosmological solutions with a nonminimally coupled scalar field},
Phys.\ Rev.\ D \textbf{61} (2000) 043506
[gr-qc/9908026].


\bibitem{Damour1992}
T.~Damour and G.~Esposito-Far\`ese,
\textit{Tensor-multi-scalar theories of gravitation},
Class.\ Quant.\ Grav.\ \textbf{9} (1992) 2093.


\bibitem{bis5}Y. Bisabr, \emph{Local Gravity Constraints and Power Law f(R) Theories},  Grav.Cosmol.\textbf{16}, 239 (2010) [arXiv:1005.5670 ]

\bibitem{Bohmer2022}
C.~G.~B\"ohmer, E.~Jensko and R.~Lazkoz,
Cosmological dynamical systems in modified gravity,
Eur. Phys. J. C \textbf{82}, 500 (2022).

\bibitem{Hrycyna2013}
O.~Hrycyna and M.~Szyd{\l}owski,
\textit{Brans--Dicke theory and the emergence of $\Lambda$CDM model},
Phys.\ Rev.\ D \textbf{88} (2013) 064018
[arXiv:1304.3300 [gr-qc]].

\bibitem{Sotiriou2010}
T.P.~Sotiriou and V.~Faraoni,
\textit{$f(R)$ theories of gravity},
Rev.\ Mod.\ Phys.\ \textbf{82} (2010) 451
[arXiv:0805.1726 [gr-qc]].

\bibitem{DeFelice2010}
A.~De Felice and S.~Tsujikawa,
\textit{$f(R)$ theories},
Living Rev.\ Relativ.\ \textbf{13} (2010) 3
[arXiv:1002.4928 [gr-qc]].

\bibitem{Alho2016}
A.~Alho, S.~Carloni and C.~Uggla,
\textit{On dynamical systems approaches and methods in $f(R)$ cosmology},
JCAP \textbf{08} (2016) 064
[arXiv:1607.05715 [gr-qc]].

\bibitem{Amendola2007}
L.~Amendola, D.~Polarski and S.~Tsujikawa,
\textit{Are $f(R)$ dark energy models cosmologically viable?},
Phys.\ Rev.\ Lett.\ \textbf{98} (2007) 131302
[astro-ph/0603703].












\end{thebibliography}
\end{document}